\documentclass[twocolumn,aps,pra,superscriptaddress,longbibliography]{revtex4-1}

\newcommand{\ket}[1]{\left|#1 \right>}

\usepackage{hyperref}
\usepackage{graphicx}
\usepackage{amsmath}
\usepackage{amssymb}
\usepackage{xcolor}
\usepackage{bm}
\usepackage{color}
\usepackage{enumitem} 
\usepackage[T1]{fontenc}
\usepackage{lipsum}
\usepackage{physics}
\usepackage{amsmath, array} 
\usepackage[normalem]{ulem}
\usepackage{soul,xcolor}
\usepackage{physics}
\usepackage{lineno}

\newcommand{\btext}[1]{{\color{black}#1}}
\usepackage[normalem]{ulem}

\usepackage[export]{adjustbox}

\begin{document}
\title{\btext{
Exploring universal and non-universal regimes of trimers from three-body interactions in one-dimensional lattices}} 

\author{Arthur Christianen} 
\affiliation{Institute for Molecules and Materials, Radboud University, Nijmegen, The Netherlands}
\affiliation{Max-Planck-Institute of Quantum Optics, Hans-Kopfermann-Stra{\ss}e 1, 85748 Garching, Germany  }

\author{John Sous}\thanks{Author to whom correspondence should be addressed. Current address: Department of Physics, Columbia University, New York, New York 10027, USA. Email: js5530@columbia.edu.}
  \affiliation{\!Department \!of \!Physics and
  Astronomy, \!University of\!  British Columbia, \!Vancouver, British
  \!Columbia,\! V6T \!1Z1,\! Canada} 
\affiliation{\!ITAMP, \!Harvard-Smithsonian \!Center \!for \!Astrophysics, \! \!Cambridge, \!Massachusetts, \!02138, \!USA}
\affiliation{\!Department \!of \!Physics, \!Harvard\!  University, \!Cambridge, \!Massachusetts, \!02138, \!USA}

\begin{abstract}
We investigate the formation of trimers in an infinite one-dimensional lattice 
model of \btext{hard-core particles} with single-particle hopping $t$ and  and \btext{nearest-neighbour two-body $U$ and 
three-body $V$} interactions of relevance to Rydberg atoms and polar molecules. 
For sufficiently attractive $U\leq-2t$ and positive $V>0$ a large trimer is 
stabilized, which persists as $V\rightarrow \infty$, while both attractive 
$U\leq0$ and $V\leq0$ bind a small trimer. The excited state 
above this small trimer is also bound and has a large extent; its behavior as $V\rightarrow -\infty$  
resembles that of the large ground-state trimer.  \btext{These large bound states appear to admit a continuum description.  Furthermore, we find that in the limit $V>>t$, $U<-2t$ the bound-state behavior qualitatively evolves with larger $|U|$ from a state described by the scattering of three far separated particles to a state of a compact dimer scattering with a single particle.}

\end{abstract}

\maketitle

\section{Introduction}
Few-body physics forms the basis of our understanding of the microscopic building units of the universe \cite{zinner:2014}. It contributes to a plethora of fundamental phenomena, including Efimov's universality \cite{efimov:1970}, quantum impurities in cold gases \cite{PolaronRev1, PolaronRev2}, quasiparticles \cite{HolPol, SSHPol} and quasiparticle pairing \cite{HolBipol,SSHBipolaron1, SSHBipolaron2, Fracton1, Fracton2} in nanoscale systems, the fractional quantum Hall effect \cite{greiter:1991}, nuclear systems \cite{brown:1969} and neutrons \cite{steiner:2012}.

A principal problem in this field is one of particles in a central potential, and the ensuing binding of bound states.  One intriguing effect prevalent in continuum systems is the formation of shallow bound states, which extend beyond the range of the potential.  Such a feeble bound state can be the lowest-energy state of the system, such as one formed in a delta-function potential in lower dimensions, or an excited state, such as Feshbach molecules \cite{Feshbachgases} and halo states \cite{Halo}.  Lattice systems with local two-body interactions do not host shallow excited bound states \cite{sawatzky:1977,PetroHubb}. It is therefore important to determine whether conditions exist under which shallow excited bound states can form in lattice systems in presence of higher-body interactions, {\em e.g.} three-body interactions.

In this work, we demonstrate that lattice systems with purely local nearest-neighbor two- and three-body interactions host bound states that extend well beyond the range of the binding forces, giving way to an emergent universality in one dimension \btext{discussed in} \cite{3bodyNishida,Ludo,1D3B2} {\btext{and}} distinct from Efimov's universality.  Namely, we demonstrate that a combination of two-site $U$ and three-site $V$ interactions stabilize universal large three-body bound states, which are either the ground state (for $V>0$) or the first excited state (for $V<0$) of the system. Tuning the strengths of interactions allows control over the size {and structure} of the bound states, providing access to the crossover between universal and non-universal few-body physics in experiments.

\section{Model}

We consider a minimal one-dimensional model of structureless fermions ({\em e.g.} spinless electrons) or hard-core bosons with nearest-neighbor (NN) hopping, and two- and three-body interactions
\begin{eqnarray}\label{model}
\hat{\mathcal{H}} = &&-t \sum_i(\hat{c}_i^\dagger \hat{c}_{i+1}+\hat{c}_{i+1}^\dagger \hat{c}_i ) + U \sum_i \hat{n}_i \hat{n}_{i+1}  \nonumber \\
&&+ V \sum_i \hat{n}_i \hat{n}_{i+1} \hat{n}_{i+2},
\end{eqnarray}
where $t$ is the hopping amplitude, $U$ is the NN two-body interaction and $V$ is the NN three-body interaction, $i$ is the site index, $\hat{c}^\dagger (\hat{c})$ is the particle creation (annihilation) operator, and $\hat{n}$ is the particle number operator. We set the lattice constant $a=1$ in what follows. This model in the NN approximation serves to provide insight into the physics of the dominant three-body interactions in a wide range of experiments.

\section{Results}

\subsection{Dimer and trimer stability}

A nonzero value of $|U|>2t$ is required to bind a dimer state, so as to compensates for the kinetic energy lost in binding \cite{sawatzky:1977, berciu:2011}. 

We study three-particle states in the infinite chain by solving the equation of motion for the Green's function $\hat{G}(\omega)=(\omega+i \eta-\hat{\mathcal{H}})^{-1}$. We derive an exact hierarchy of equations of motion for three-particle propagators $G(m_1,m_2;n_1,n_2;K,\omega)=\langle K,m_1,m_2|\hat{G}(\omega)|K,n_1,n_2 \rangle$ defined for states $|K,n_1,n_2 \rangle =\frac{1}{\sqrt{N}}\sum_i e^{iKR_i} \hat{c}_{i-n1}^\dagger \hat{c}_i ^\dagger \hat{c}_{i+n2}^\dagger |0 \rangle$ \cite{berciu:2011}.  A stable attractively (repulsively) bound trimer (also known as trion) corresponds to the appearance of a discrete pole in the Green's function below (above) the continuum of scattering states. 

To identify stable trimers we search for discrete peaks outside of the three-particle continuum. This consists of scattering states of three free particles, $1+1+1$, and those of a dimer and a free particle, $2+1$.  In the current work, we discuss trimers formed below the continuum ($U/t<0$), {\em i.e.} attractively bound trimers.

\begin{figure}[t]
\includegraphics[width=0.98\columnwidth,left]{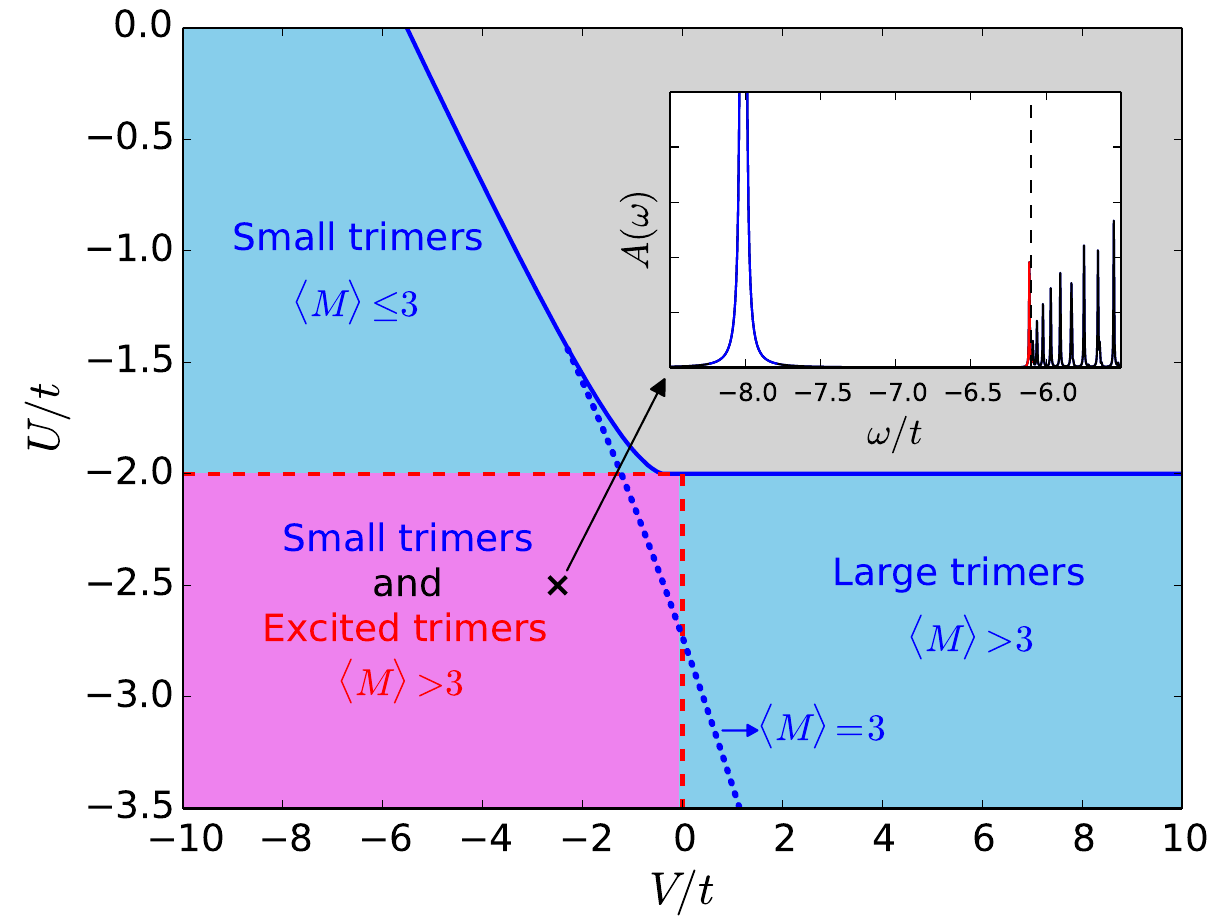}
\caption{(color online) Trimer stability diagram at $K=0$ as a function of $V$ and $U$ in units of $t$.  The grey area indicates the continuum. The solid blue line identifies the boundary of the stability region of attractively bound states, while dashed red lines identify regions where an excited-state (ES) trimer co-exists with the ground-state (GS) trimer. The crossover from small ($\langle M \rangle \leq 3$) to large ($\langle M \rangle >3$) GS trimers is indicated by the dotted blue line.  In the inset, we show the spectral function $A(\omega)=-\frac{1}{\pi} \Im G(1,1;1,1;0,\omega)$ for the parameter values indicated by the cross: $V = -2.5t$, $U=-2.5t$, demonstrating the appearance of the GS (blue) and ES (red) trimer peaks below the edge of the continuum (dashed line).
}
\label{fig1}
\end{figure}

In Figure \ref{fig1}, we plot the stability diagram for bound states with total quasimomentum $K=k_1+k_2+k_3=0$. The solid blue line identifies the stability behavior of attractive trimers. To characterize different regimes of physical behavior, we compute the average size of the trimer $\langle M\rangle$, where $M=n_1+n_2$ is the distance between the two outer particles in a given configuration of the trimer.

First consider the upper-left quadrant of the diagram ($V\leq0$, $U\geq -2t$). For $U=0$, a bound trimer (blue upper-left region of Figure \ref{fig1}) appears for $V\lessapprox-5.5t$ with particles tightly bound in the trimer state $\langle M \rangle \leq 3$ as expected of the short-range three-body attraction. Increasingly attractive $U$ values lead to more tightly bound trimers and naturally lowers the $V$ needed for binding.

Now consider the lower-right quadrant ($U\leq-2t$, $V\geq 0$). Surprisingly, for sufficiently attractive $U \leq -2t$, trimers are always stable regardless of the magnitude of the repulsive $V$. This behavior persists for extremely large $V$ (not shown). The large $V$ effectively pushes the particles in the trimer apart as it becomes energetically costly to occupy three consecutive sites, but fails to completely break down the trimer. These exotic large trimers with $\langle M \rangle >3$ are bound by non-perturbative higher-order interactions.   

We now discuss the lower-left quadrant of Figure \ref{fig1}, ($U\leq-2t$, $V\leq 0$). As expected, these strongly attractive $U$ and $V$ bind a small trimer. Interestingly, however, a second bound state appears below the continuum (red region of Figure \ref{fig1}), see also the inset. These feebly bound excited-state (ES) trimers are extended ($\langle M \rangle >3$) similar to the ground-state (GS) trimers at large repulsive $V$.

\btext{The large trimer states extend beyond the lattice scale pointing to an emergent long-wavelength continuum description insensitive to microscopic details, discussed below.  
}

We note in passing that, for $K = 0$, only small, and no large, repulsively bound trimers appear above the continuum (not shown).

\begin{figure}[t]
(a) \newline
\centering
 {\includegraphics[width=0.98\columnwidth]{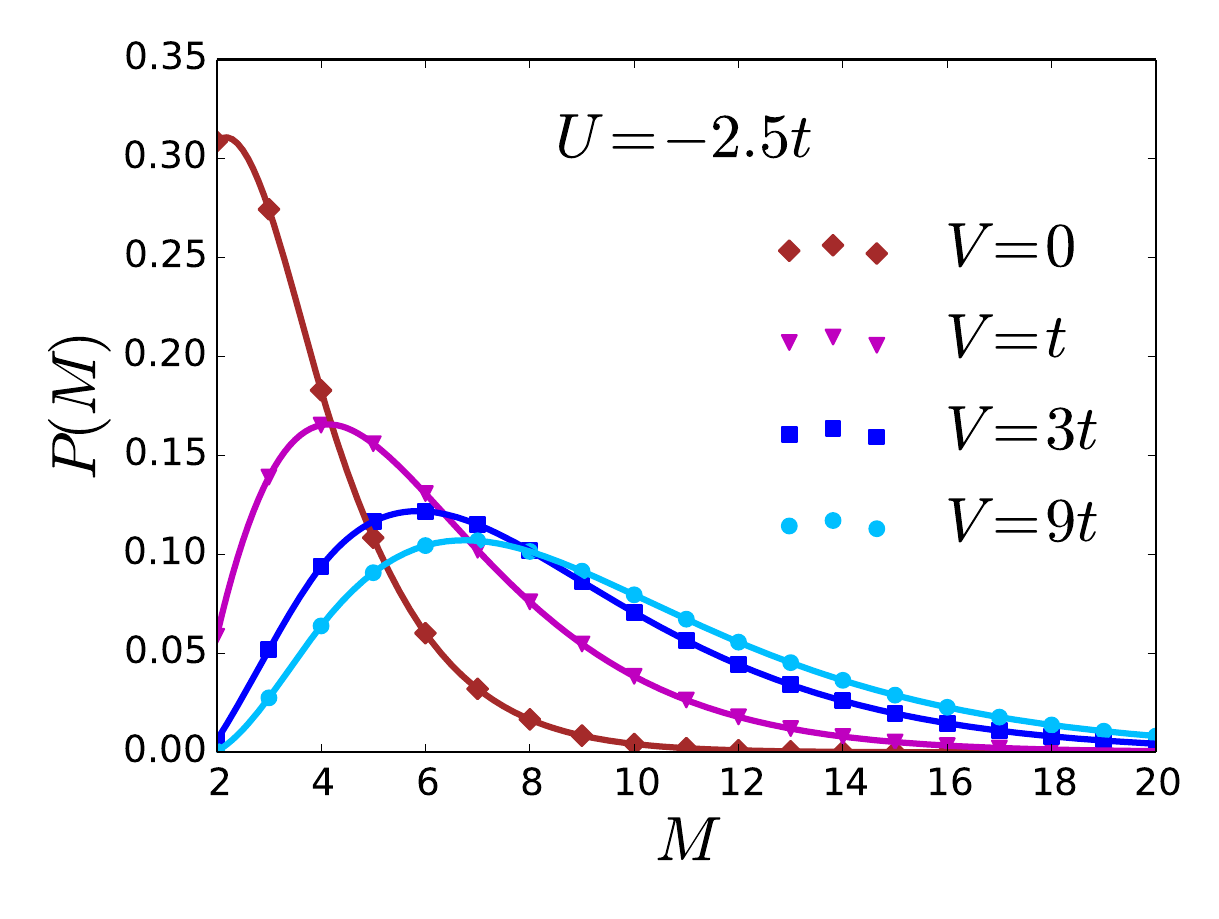}} 

(b) \newline
\centering
{\includegraphics[width=0.9824\columnwidth]{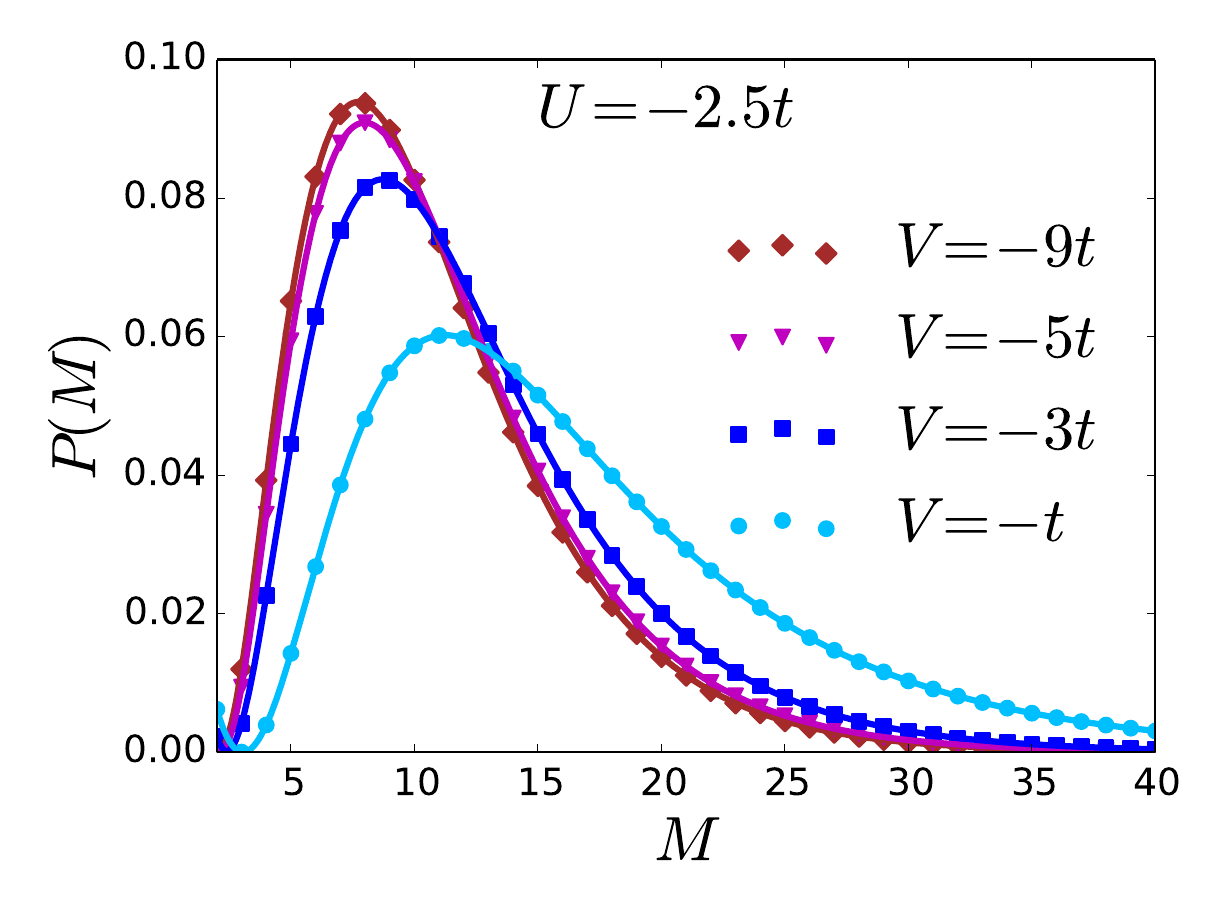}} 
\caption{(color online) Analysis of the size of trimers:
The probability $P(M)=\sum_{n_1+n_2=M} |\langle 0,n_1,n_2|0,\alpha_{\mathcal{T}} \rangle|^2$ for the two outer particles in a trimer to be $M=n_1+n_2$ sites apart at $U=-2.5t$ and various values of $V$ for the (a) ground-state (GS) trimers (lower-right quadrant of Figure \ref{fig1}) and (b) excited-state (ES) trimers (lower-left quadrant of Figure \ref{fig1}). The two trimers exhibit qualitatively similar behavior with increasing V (compare lines of the same colors and symbols in (a) and (b)). Further analysis of the $M=8$ component of the GS trimer is presented in Figure \ref{fig7}.}
\label{fig2}
\end{figure}

\subsection{Trimer structure}

To shed light on the mechanism behind the formation of trimers and their structure, we analyze the probability density 
\begin{equation}
P(M)=\sum_{n_1+n_2=M} |\langle 0,n_1,n_2|0,\alpha_{\mathcal{T}} \rangle|^2
\end{equation}
of the trimer eigenstates $\ket{0,\alpha_{\mathcal{T}}}$ at quasimomentum $K=0$. \btext{(In what follows we focus exclusively on $K=0$ trimers.)}
 
We study $P(M)$ as a function of $V$ for a fixed $U=-2.5t$ in Figure \ref{fig2} for (a) the GS trimers and (b) the ES trimers. 

\begin{figure}[t]
{\includegraphics[width=0.9824\columnwidth]{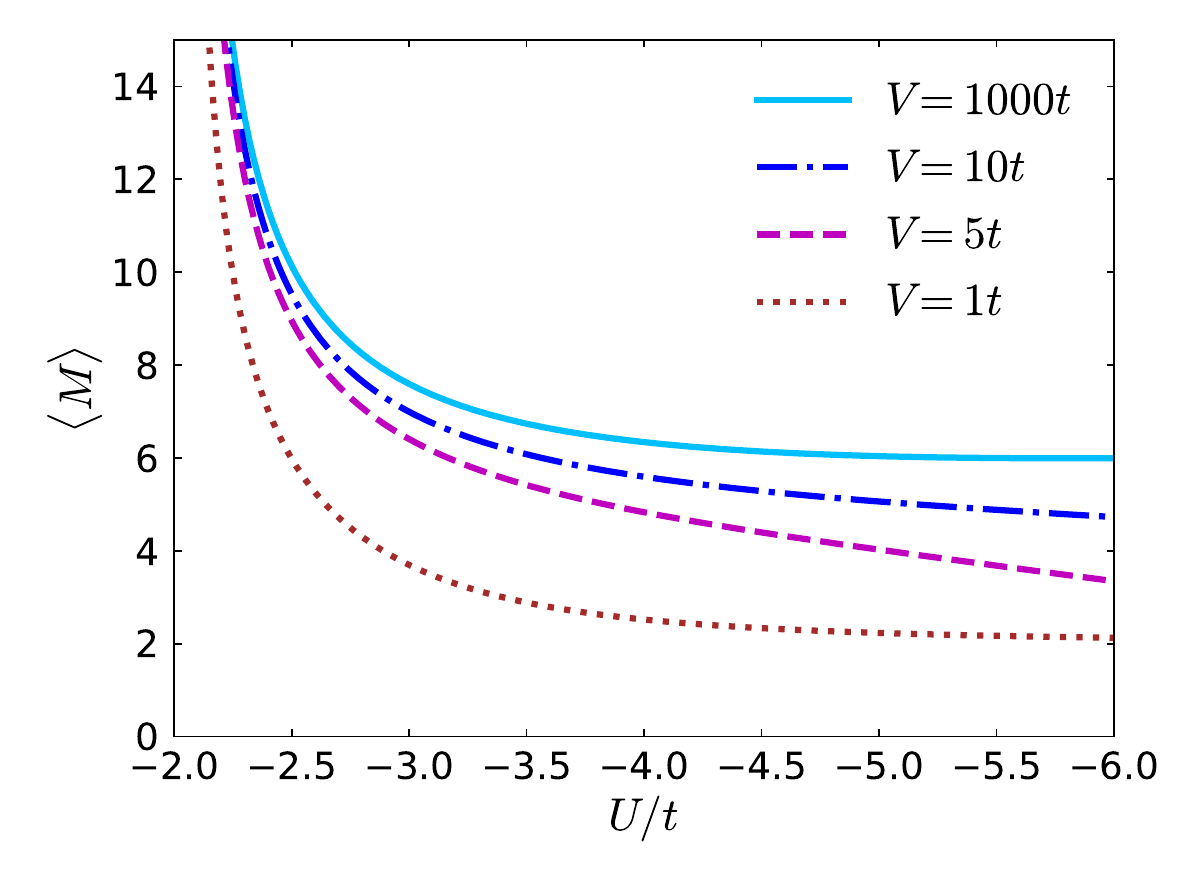}} 
\caption{(color online) The expectation value of the GS trimer size $\langle M\rangle$ as a function of $U$ for several values of $V$. The size of the trimer diverges as $U$ approaches $-2t$ for all $V\geq0$, implying emergent universal behavior.}
\label{fig3}
\end{figure}

The size of the GS trimer evolves with $V$ from small ($\langle M \rangle \approx 3$) to large ($\langle M \rangle > 3$) (Figure \ref{fig2}(a)), see also the dotted line in Figure \ref{fig1}. This crossover behavior is characterized by a shift in the maximum of $P(M)$ to larger values. In comparison, the ES trimer is much more extended, however its spread also grows with $V$ (Figure \ref{fig2}(b)). \btext{That these bound states extend over large distances (Figure \ref{fig3})} is an indication \btext{that their physics is amenable to a continuum description insensitive to the lattice details with few universal parameters (two-body $a_2$ and three-body $a_3$ scattering lengths) as developed in \cite{3bodyNishida,Ludo,1D3B2}.}

\begin{figure}[t]
\includegraphics[width=\columnwidth,left]{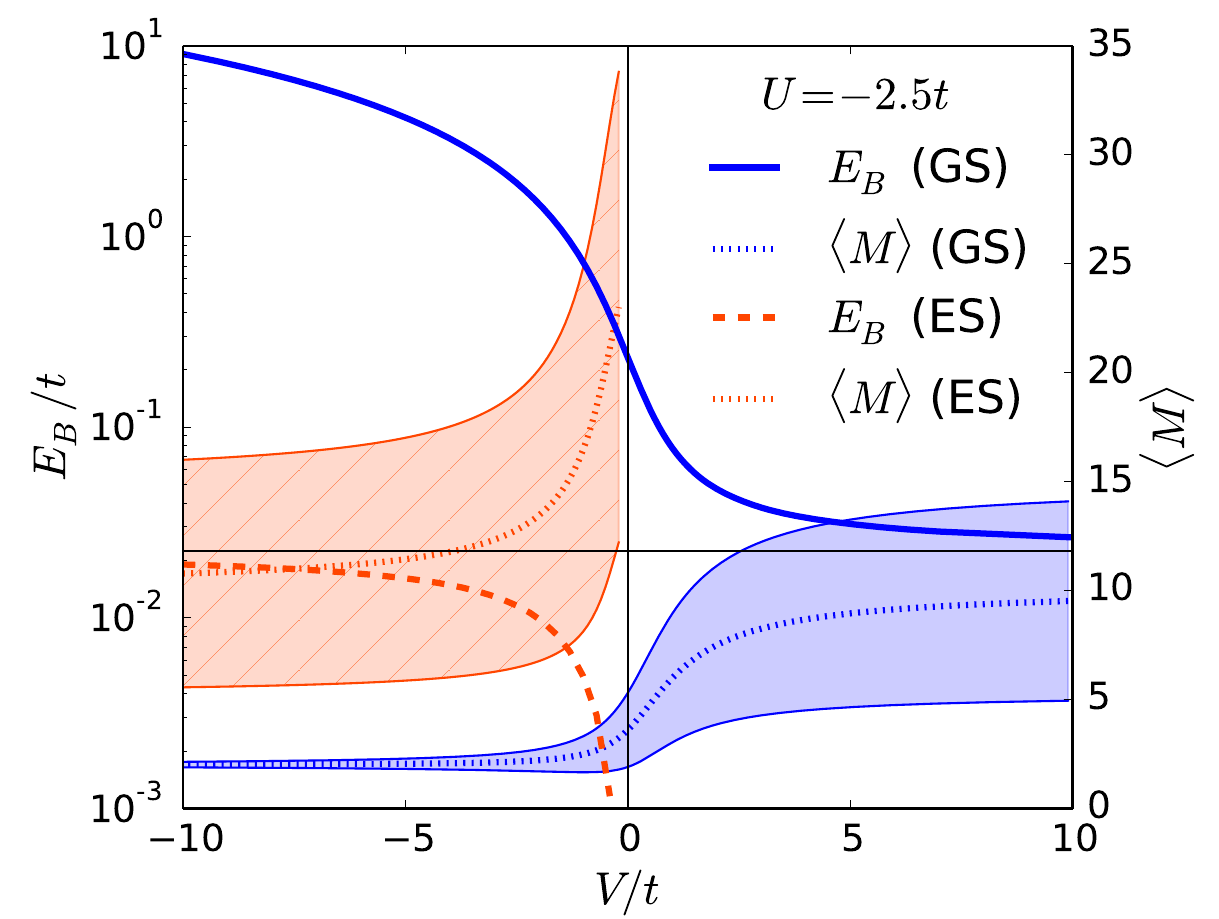}
\caption{(color online) Binding of ground-state (GS) (blue) and excited-state (ES) (salmon) trimers at $U=-2.5t$ as a function of $V$. We plot the binding energy $E_B$ (solid and dashed lines) and the average trimer size $\langle M \rangle$ (dotted lines) with $\langle M \rangle \pm \sigma$ (boundary of the shaded regions), where $\sigma$ is the standard deviation of $P(M)$. $E_B$ approaches the horizontal black line in the asymptotic limit $V\rightarrow -\infty (\infty)$ for the ES (GS) trimer.}
\label{fig4}
\end{figure}

To corroborate this picture, we study the binding energy $E_B$ of the trimer bound states. In Figure \ref{fig4}, we plot $E_B$ along with $\langle M \rangle$ and its spread for the GS (blue) and ES (salmon) trimers as a function of $V$ for an exemplary $U=-2.5t$. As expected, for $V<0$, $E_B$ (solid line) of the GS trimer grows with $\abs{V}$, saturating at the smallest possible size of $M= 2$ with essentially no spread. For repulsive $V>0$, the binding energy decreases, asymptotically approaching $E_B \approx 0.0225t$ (horizontal solid line), and both $\langle M \rangle$ and its spread increase, saturating at $\langle M \rangle \approx 10.14$. Intriguingly, we find the same asymptotic behavior for the ES trimer as $V \rightarrow -\infty$ (we have verified this numerically).  \btext{From this, we see that the universal trimers behave with no sensitivity to $V$ in the large $|V|>>t$ limits; the trimer's structure and binding energy are nearly unaffected by variations in $V$, and thus its behavior can be analyzed by considering a continuum description insensitive to the microscopic parameters.}

We can understand this behavior as follows. In the limit $V \rightarrow -\infty$, the ground-state trimer $|\Psi_{\rm GS}\rangle$ asymptotically approaches the state with the smallest possible size and no spread, {\em i.e.} $|K,1,1\rangle$.  The ES trimer must be orthogonal to the GS trimer, and in this limit we find $\langle \Psi_{\rm ES} | \Psi_{\rm GS}\rangle \rightarrow \langle \Psi_{\rm ES} |K,1,1\rangle =0$. On the other hand, in the limit $V \rightarrow \infty$, the NN configuration $|K,1,1\rangle$ in the trimer wavefunction is energetically forbidden. This reflects in the relation $\langle \Psi_{\rm GS}|K,1,1\rangle =0$.  The problem of finding the Hamiltonian spectrum requires diagonalizing the Hamiltonian operator whose structure then takes the same exact form in these two asymptotic limits, explaining the resemblance between the asymptotic forms of the ES and GS trimers. \btext{This asymptotic correspondence extends to the continuum as was shown in \cite{3bodyNishida}, where the energy of the deepest bound-state in the limit $a_3 \rightarrow \infty$ was demonstrated to match that of an excited state in the limit $a_3 \rightarrow 0$.} {\btext{In the present work,} we find that the large GS and ES trimers asymptotically behave as $E_B \rightarrow E_0(U)  \exp\Big(\gamma(U) t/V\Big)$, where $E_0(-2.5t)  \approx 0.0225t$ and $\gamma(-2.5t) \approx 0.5 \pi$ \footnote{We write $\gamma(U)$ as a multiplicative factor of $\pi$ to emphasize its geometric origin that results from regularizing the divergent momentum integration in a continuum theory, see Refs. \cite{1DDrop,3bodyNishida}. A precise $\gamma(-2.5t) = 0.527$ results in a fit within an accuracy of $10^{-7}$ in the range of $|V| \in [100t,1000t]$.  In this range \unexpanded{$\langle M \rangle$} and its spread vary less than a percent.}. {While we cannot determine $a_2$ and $a_3$ from the microscopic parameters unambiguously, we view this exponential dependence of $E_B$ on $1/V$ (or equivalently, the inverse logarithmic dependence of $V$ on $E_B$) as suggestive of the signature of three-body universality in one dimension.\cite{3bodyNishida,Ludo,1D3B2,1DDrop}.}

\subsection{Behavior of large trimers: Universal regimes}

 \begin{figure}[t]
\includegraphics[width=\columnwidth,left]{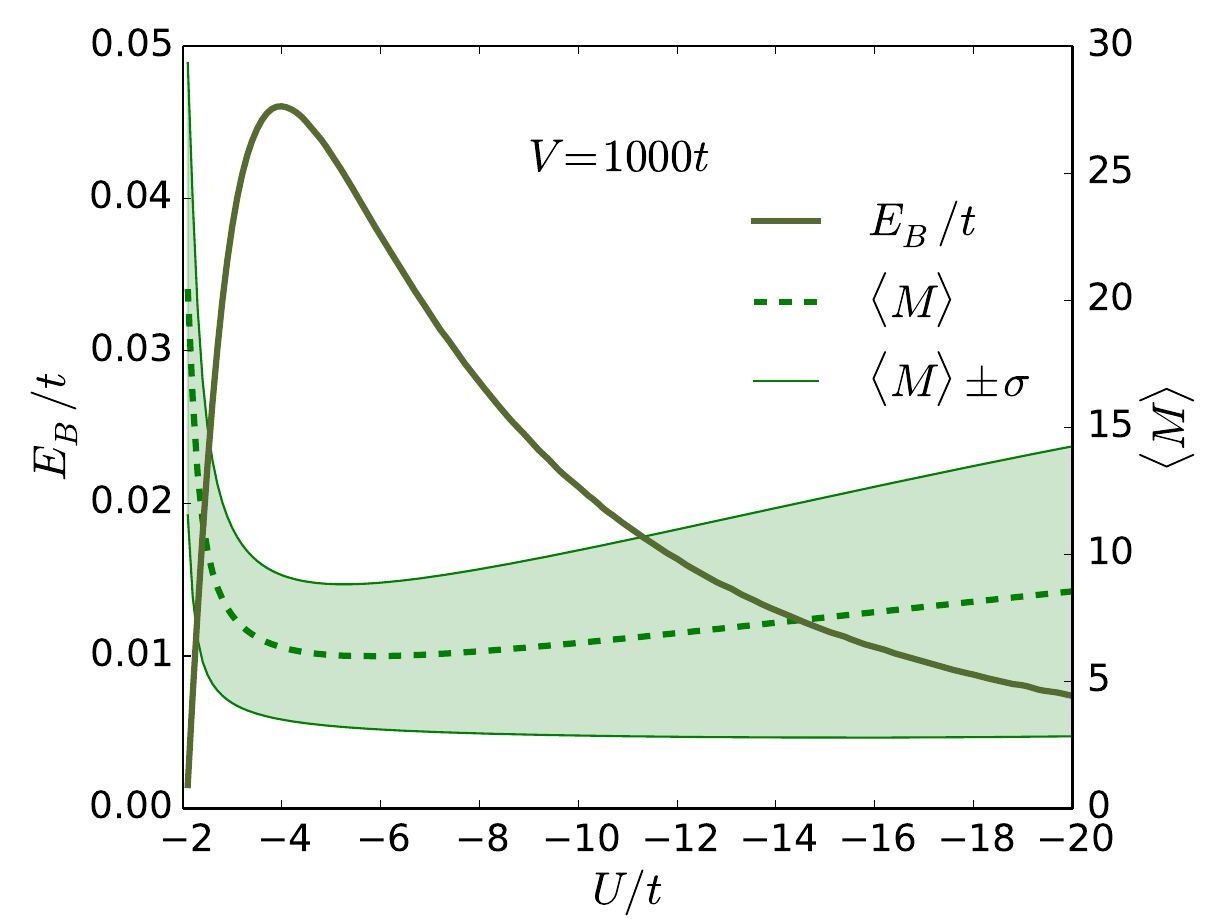}
\caption{(color online) Binding mechanism of the large GS trimer at $V=1000t$ as a function of $U$. We plot the binding energy $E_B$ (solid green line), the average trimer size $\langle M \rangle$ (dashed green line), and $\langle M \rangle \pm \sigma$ (boundary of shaded regions), where $\sigma$ is the standard deviation of $P(M)$. The shaded region shows the spread of $P(M)$ about the average $\langle M \rangle$.}
\label{fig5}
\end{figure}

\begin{figure}[t]
(a) \newline
\centering
 {\includegraphics[width=1.0\columnwidth]{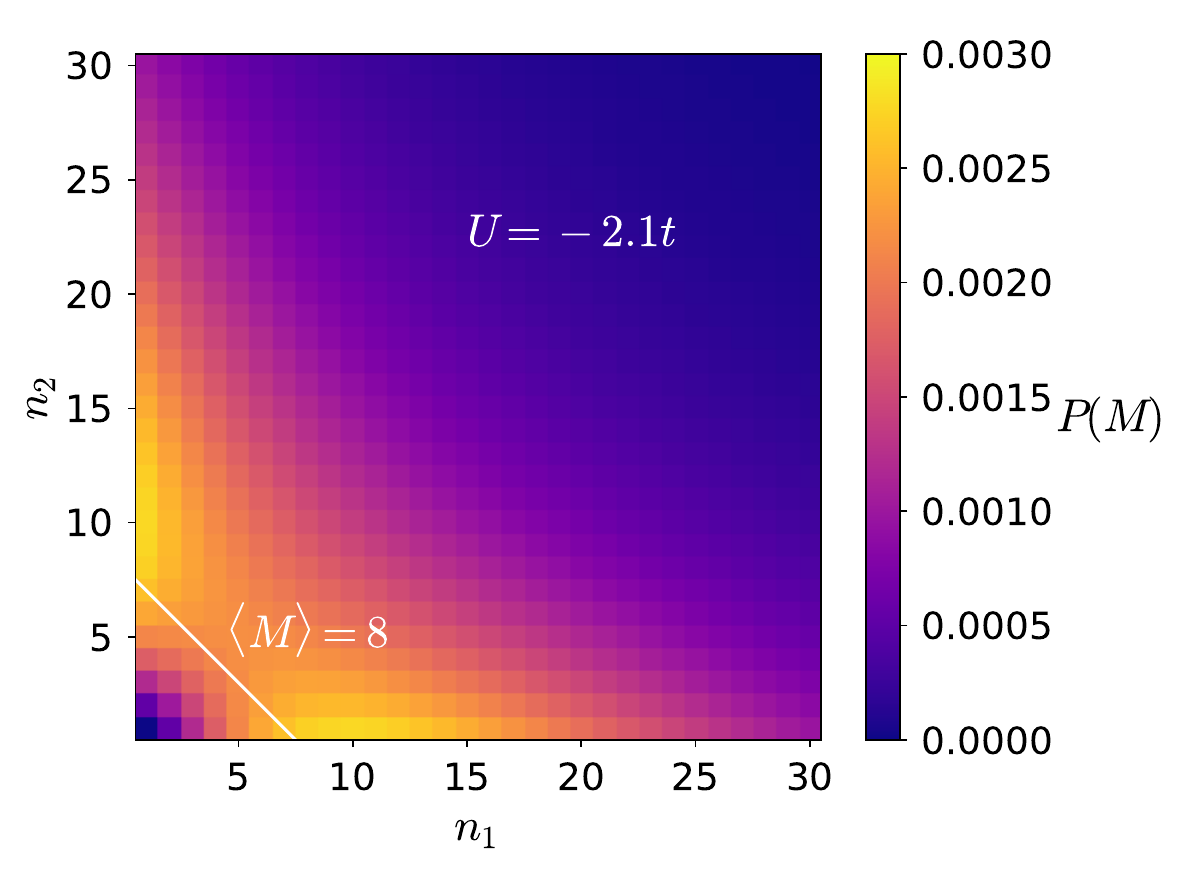}} 
\centering
(b) \newline
\vspace{-0.6cm}
\centering
\hspace{-0.7cm}{\includegraphics[width=1\columnwidth]{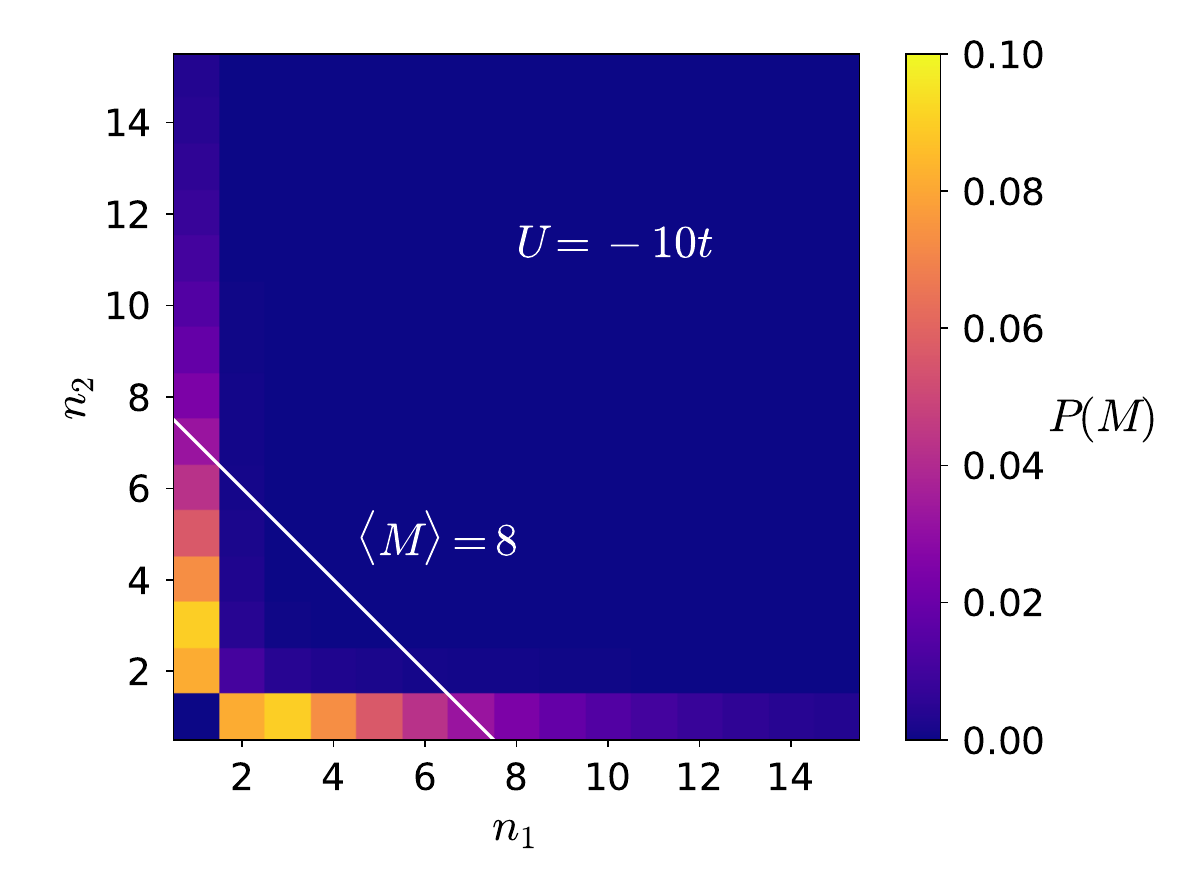}} 
\caption{(color online) Spatial structure of GS trimer shown as a color plot of the probability $P(M)=\sum_{n_1+n_2=M} |\langle 0,n_1,n_2|0,\alpha_{\mathcal{T}} \rangle|^2$ for the two outer particles in a trimer to be $M=n_1+n_2$ sites apart for $V=1000t$ at (a) $U=-2.1t$  and  (b) $U=-10t$. Analysis of the trimer's $M=8$ component (white line) is presented in Figure \ref{fig7}.
}
\label{fig6}
\end{figure}

{We now turn to the large trimers ($V>>t$,  $U<-2t$). A long-wavelength description is only expected to work when the relevant scattering lengths are much larger than the lattice spacing, i.e., when the trimers become extended. When $|U|>>2t$ this may no longer be the case, as the three particles in the trimer become closer. The correspondence between the GS and the ES trimers, however, continues to hold for any value of $U<-2t$.  This suggests that a robust or stronger type of universality may be at play, as explained below.}

In Figure \ref{fig5} we analyze $E_{B}$ along with the corresponding $\langle M \rangle$ of the GS trimer for $V=1000~t$ at $K=0$ as a function of $U$. \btext{(See Figure \ref{fig6} for a spatial visualization of the trimer's wavefunction.)} As expected $E_{B}$ increases with increasingly attractive $U$, but only up to $U\sim-3.9t$. At this $U$, $E_{B}$ develops a maximum followed by a rapid decrease. This striking behavior accompanies an opposite trend in $\langle M \rangle$ which has a minimum roughly coinciding with the maximum in $E_B$.

\begin{figure}[t]
\centering
\includegraphics[width=\columnwidth,left]{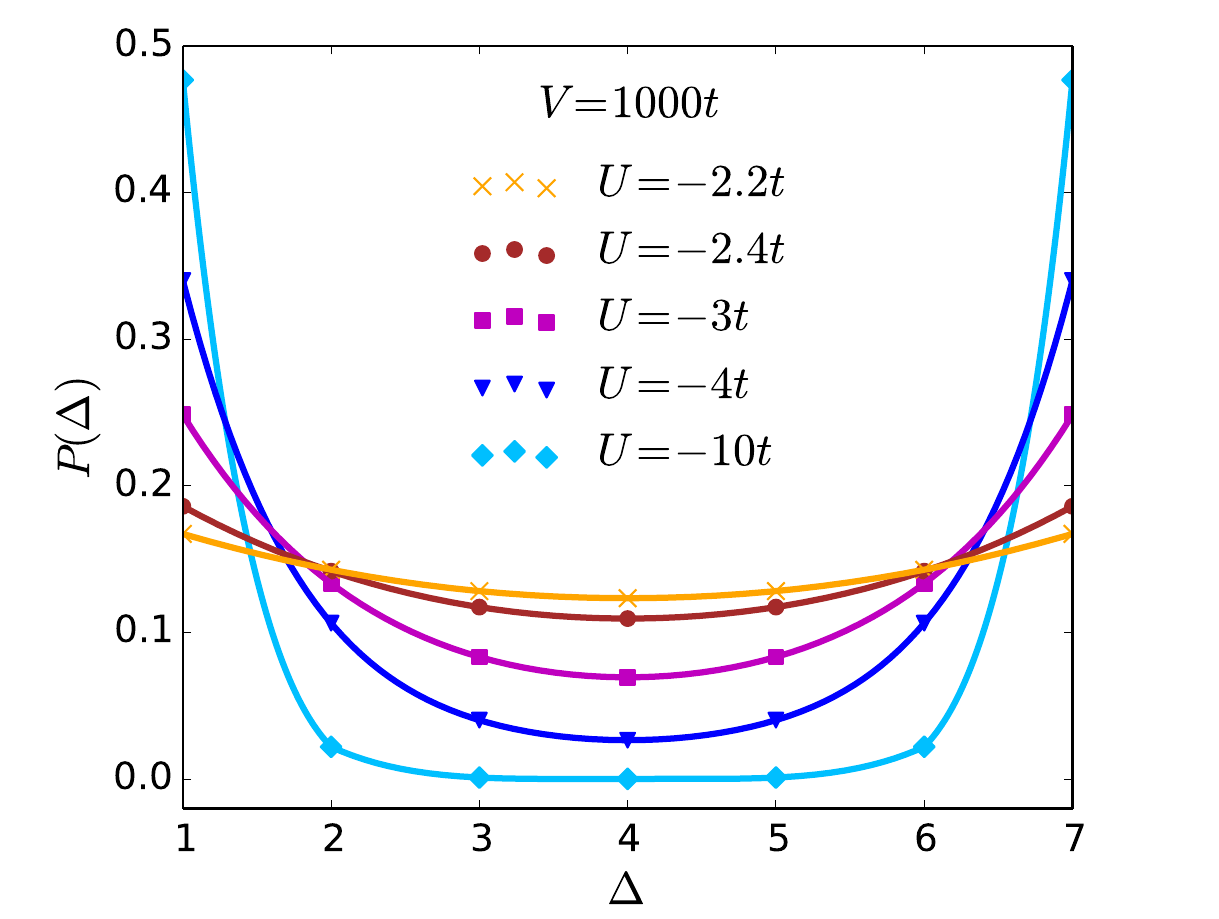}
\caption{(color online) Analysis of the internal structure of the large GS trimer through the probability $P(\Delta)= \frac{|\langle 0,\Delta,M-\Delta|0,\alpha_{\mathcal{T}} \rangle|^2}{P(M)}$ to find the central particle $\Delta$ sites apart from the outermost left particle for the $M = 8$ component of the trimer wavefunction at $V = 1000t$ and for different values of $U$.}
\label{fig7}
\end{figure}

To explain this maximum in the binding energy as a function of $U$, we consider the probability density
\begin{equation}
P(\Delta)= \frac{|\langle 0,\Delta,M-\Delta|0,\alpha_{\mathcal{T}} \rangle|^2}{P(M)}
\end{equation}
to find the central particle at a distance $\Delta$ from the outer left particle in the trimer for a given $M$ component of the wavefunction. In Figure \ref{fig7}, we plot $P(\Delta)$ for the $M=8$ component of the GS trimer wavefunction at a fixed $V=1000t$ for different values of $U$. Simple perturbative arguments suggest that binding should be facilitated by the formation of configurations with NN particles $\Delta=1,7$ as a result of the attractive NN two-body interaction. \btext{In contrast, for small $\left|U\right|$ we find that the central particle is only slightly more likely to be NN to either outer one and has a large probability to be anywhere in between.} \btext{Ultimately, a larger} attractive $U$ favors NN configurations with $\Delta=1,7$.

With this insight in hand we can qualitatively explain the behaviour in Figure \ref{fig5}.  In the region of moderate $U>-3.9t$, an increasing $|U|$ favors a smaller trimer as intuitively expected, and thus $E_B$ grows till it approaches a maximum.} Larger $U<-3.9t$, however, forces the trimer into configurations with two NN particles and the third further apart ({\em e.g.} $U=-10t$ result in Figure \ref{fig7}) accompanied by an increase in $\langle M \rangle$, and thus $E_B$ decreases. This trimer configuration is a weakly bound state of a strongly bound dimer and a single particle. $\langle M \rangle \pm \sigma$ (shaded region of Figure \ref{fig5}), where $\sigma$ is the standard deviation of $P(M)$, shows larger spread for more attractive $U$ corroborating this picture of a dimer and a loosely bound third particle. 

These results point to a non-perturbative binding mechanism: The large timers are bound by higher-order interactions that mediate long-range binding yet avoid the forbidden $M=2$ configuration. Furthermore, this pattern of decrease in $E_B$ for large trimers composed of NN pairs and a loosely bound particle indicates that configurations with the central particle `free' in between the outer two play a crucial role in binding. There, the central particle mediates a three-body force through pairwise interactions with the outer two. This is most efficient in configurations with the central particle close to both the outer two, a situation favorable in smaller trimers formed for modest $U$. Larger $U$ forces the central particle closer to one of the outer two, ultimately weakening the binding to the other one, which leads to a larger trimer with a $2+1$-like structure.

\btext{The increase of $\langle M \rangle$ as a function of $|U|$ suggests that besides the continuum description for small $\delta_U = U-2t$ in terms of the scattering of three particles, a distinct theory in terms of the scattering of a dimer and a particle may be relevant in the limit of strong attractive $U$. This evolution with attractive $U$ into qualitatively different long-wavelength behavior shows the subtleties associated with obtaining a continuum description of the low-energy physics of one-dimensional trimers in a lattice realization.}

\section{Conclusion}
We studied the interplay of two- and three-body interactions in a minimal one-dimensional lattice model. We constructed three-body bound-state stability diagram identifying regions in parameter space of attractively bound trimers. Trimers form even in the limit of infinite three-body repulsion. An ES bound trimer appears for attractive $V$ and persists as $V \rightarrow - \infty$, where it develops asymptotic behavior similar to that of the GS trimer as $V \rightarrow \infty$.

These large trimers are bound by non-perturbative long-range forces mediated by short-range interactions, which favor large configurations with the central particle free in between the outer two. They extend over several lattice spacings pointing to emergent long-wavelength universality, and are thus of great interest to efforts targeting the creation of large coherent quantum objects with non-trivial internal structure.

Our analysis applies to few-body bound states realized, for example, with polar molecules in optical lattices \cite{buchler:2007} or Rydberg atoms in tweezers \cite{bernien:2017}, and to systems with three-site blockade ($V\rightarrow\infty$ limit), such as Coulomb blockaded Rydberg gases \cite{RydbergBlockade} and quantum dots \cite{CBlockade}. Other potential experimental systems with few-body interactions include trapped ultracold gases \cite{dipolarDemler, TrimerZoller,petrov:2014}, ultracold atoms in optical lattices \cite{daley:2009,johnson:2009,mazza:2010,Daley23body,PetrovLattice}, Rydberg excitations in cold gases \cite{HRreview,greene:2000,RydMolP,Rydberg2,RydbergPolaron,fey2018observation,RydFermiSous}, Rydberg slow light polaritons \cite{jachymski:2016, gullans:2016, Lukin3exp, stiesdal2018observation}, ion traps \cite{Ion3BodySpain}, optics coupled-cavity arrays \cite{prasad2017effective}, and circuit QED systems \cite{CircuitQED}, where many of the ideas we discuss and others \footnote{For example, see Refs. \cite{3bodysoftcore,TrionHofs,UniversalClusters, QuantumAnamoly}.} can be investigated.  \btext{We note the model Hamiltonian we use serves to adequately describe interacting dipolar systems only in the weakly interacting limit, since these may not assume a lattice Hubbard-like description in the strongly interacting limit \cite{RefB2, RefB1, RefB3}.}
Our method accurately simulates spectroscopy in the frequency domain (inset of Figure \ref{fig1}) and can be extended to analyze the time-resolved response in one and higher dimensions.

Our results imply universality for fermionic trimers in one dimension. An interesting question arises whether statistics play a role in universality in one dimension when the equivalence between hard-core bosons and spinless fermions \footnote{For a discussion of hard-core and Fermi statistics in one dimension see Refs. \cite{Stat1,Stat2,Stat3}} breaks down, {\em e.g.} for soft-core interactions. Another emergent line of inquiry is whether the universal correspondence between GS and ES complexes persists for larger number of particles.

\begin{acknowledgments} 
We acknowledge helpful discussions with Roman Krems, Richard Schmidt, Mona Berciu, Kun Chen, Sergej Moroz and Ed Grant. A.~C. acknowledges support from the Radboud Honours Academy and the hospitality of the University of British Columbia. J.~S. acknowledges support from the Natural Sciences and Engineering Research Council of Canada (NSERC), a visiting student fellowship at the Institute for Theoretical Atomic, Molecular, and Optical Physics (ITAMP) at Harvard University and the Smithsonian Astrophysical Observatory, and the hospitality of the Stewart Blusson Quantum Matter Institute at the University of British Columbia.
\end{acknowledgments}

\Urlmuskip=0mu plus 1mu\relax

%

\end{document}